\documentclass[sigconf, 9pt, nonacm]{acmart}
\usepackage[version=4]{mhchem} 
\usepackage{caption}
\usepackage{subcaption}
\usepackage{colortbl}
\usepackage[normalem]{ulem}
\usepackage{xcolor}
\usepackage{afterpage}
\usepackage{float}

\setcopyright{none}
\renewcommand\footnotetextcopyrightpermission[1]{} 
\newcommand{\revision}[1]{\textcolor{black}{#1}}

\AtBeginDocument{%
  \providecommand\BibTeX{{%
    \normalfont B\kern-0.5em{\scshape i\kern-0.25em b}\kern-0.8em\TeX}}}

\acmConference[EWSN]{EMBEDDED WIRELESS SYSTEMS AND NETWORKS}{10-13, December 2024}{Abudabi,UAE}
\begin{document}

\title{Feasibility of Radio Frequency Based Wireless Sensing of Lead Contamination in Soil}


\author{Yixuan Gao}
\email{yixuan@cs.cornell.edu}
\orcid{0000-0003-1778-3104}
\affiliation{%
  \institution{Cornell Tech}
  \streetaddress{2 W Loop Road}
  \state{New York}
  \country{USA}
  \postcode{10044}
}

\author{Tanvir Ahmed}
\email{tanvir@infosci.cornell.edu}
\orcid{0000-0002-9468-5033}
\affiliation{%
  \institution{Cornell Tech}
  \streetaddress{2 W Loop Road}
  \state{New York}
  \country{USA}
  \postcode{10044}
}

\author{Mikhail Mohammed}
\email{mikhail.mohammed05@bcmail.cuny.edu}
\affiliation{%
  \institution{Brooklyn College of the City University of New York}
  \state{New York}
  \country{USA}
}

\author{Zhongqi Cheng}
\email{ZCheng@brooklyn.cuny.edu}
\affiliation{%
  \institution{Brooklyn College of the City University of New York}
    \state{New York}
  \country{USA}
}

\author{Rajalakshmi Nandakumar}
\email{rajalakshmi.nandakumar@cornell.edu}
\affiliation{%
  \institution{Cornell Tech}
  \streetaddress{2 W Loop Road}
  \state{New York}
  \country{USA}
  \postcode{10044}
}


\renewcommand{\shortauthors}{Gao, et al.}

\begin{abstract}

Widespread \ce{Pb} (lead) contamination of urban soil significantly impacts food safety and public health and hinders city greening efforts. However, most existing technologies for measuring \ce{Pb} are labor-intensive and costly. In this study, we propose SoilScanner, a radio frequency-based wireless system that can detect \ce{Pb} in soils. This is based on our discovery that the propagation of different frequency band radio signals is affected differently by different salts such as \ce{NaCl} and \ce{Pb(NO_3)_2} in the soil. In a controlled experiment, manually adding \ce{NaCl} and \ce{Pb(NO_3)_2} in clean soil, we demonstrated that different salts reflected signals at different frequencies in distinct patterns. In addition, we confirmed the finding using uncontrolled field samples with a machine learning model. Our experiment results show that SoilScanner can classify soil samples into low-Pb and high-Pb categories (threshold at 200 ppm) with an accuracy of 72\%, with no sample with $>500$ ppm of Pb being misclassified. The results of this study show that it is feasible to build portable and affordable \ce{Pb} detection and screening devices based on wireless technology.

\end{abstract}

%
%




\settopmatter{printfolios=true}

\maketitle


\begin{table*}
    \caption{Related work of RF-based soil sensing.}
    \label{tab:relatedwork}
    \begin{tabular}{|l|l|l|l|
    >{\columncolor[HTML]{FD6864}}l |l|}
    \hline
    \textbf{System}                                           & \textbf{Method}        & \textbf{Moisture} & \textbf{Salinity} & \textbf{Lead} & \textbf{Contribution Sumary}                           \\ \hline\hline
    UHF RFID tags \cite{aroca2016application} & RFID - signal strength & Y                 & N                 & N                           & First COTS system testing soil moisture                \\ \hline
    Strobe \cite{ding2019towards}             & Wifi - relative ToF    & Y                 & Y                 & N                           & First work considering salinity of soil \\ \hline
    GreenTag  \cite{wang2020soil}             & RFID - signal strength & Y                 & N                 & N                           & Greenhouse setting                                     \\ \hline
    CoMEt  \cite{khan2022estimating}          & RF - ToF 2GHz-4GHz     & Y                 & N                 & N                           & Moisture estimation without burying parts                                     \\ \hline
    SoilScanner{[}This Work{]}                      & Wifi + RFID            & N/A               & N/A               & \cellcolor[HTML]{67FD9A}Y   & First work considering individual salt in soil         \\ \hline
    \end{tabular}
\end{table*}

\section{Introduction}

Urban soils are significant resources and provide essential ecological services such as growing produce, assimilation of organic waste, stormwater management, greening cities, improving air and water quality, and combating urban heat island effects \cite{cheng2021urban}. However, numerous studies have shown that urban soils are often contaminated, primarily due to historical and current anthropogenic activities \cite{qian2017risk}. 

Lead (\ce{Pb}), an invisible, odorless neurotoxin, is of particular concern, given its widespread presence in the environment and strong association with neurocognitive disorders and aggression in adolescents - especially for children \cite{ara2015lead, mitra2017clinical}. \ce{Pb} is found to be present at elevated levels in urban soils worldwide. In New York City, it was found that over $50\%$ of the garden soils tested contained more than 400 \revision{parts per million (}ppm\revision{)} of \ce{Pb} – the previous general threshold set by the U.S. Environmental Protection Agency (EPA) and the New York State Department of Environmental Conservation \cite{cheng2015trace}. \revision{In an official statement released on January 17, 2024, to strengthen the safeguards to protect families and children from Pb-contaminated soil, the U.S. EPA lowered the screening level for \ce{Pb} in soil at residential properties from 400ppm to 200ppm \cite{epa2024lead}, which is now the current screening standard.} There is an urgent need to screen urban soils for traces of metal contaminants (such as \ce{Pb}) as they can significantly impact public health and the safety of food grown in urban community gardens \cite{malavolti2020lead, nag2022human}. 

Currently, composite soil samples are commonly sent to commercial or academic laboratories for analysis, utilizing chemical processing techniques and advanced instrumentation. Such analysis tends to require substantial labor and incur significant expenses \cite{soodan2014analytical}, and thus poses challenges for many urban communities. These communities, often characterized by economic disadvantages, a prevalence of minority or marginalized populations, and a disproportionate burden of environmental contamination, face particular difficulties in accessing related resources. Furthermore, soil \ce{Pb} is highly heterogeneous at a small scale, even within the same garden \cite{juhasz2013required, gonzales2021agreement}. \ce{Pb} levels can vary by more than an order of magnitude at different locations within the same garden \cite{bechet2018spatial}. Thus, a composite soil sample is not able to reveal such variations and will miss hotspots that may pose the most health risks. While Portable X-Ray Fluorescence (pXRF\cite{WEINDORF20141}) has emerged as a handy tool for in-situ (or lab) screening of \ce{Pb} and other metals in soils \cite{paltseva2022prediction, gholizadeh2015estimation, schwartz2011reflectance, song2012diffuse, wu2007mechanism, liu2019estimation, sun2017estimating}, the instrument typically costs \$20,000-60,000, which is not affordable for most communities. Therefore, there is a need to develop low-cost alternatives to detect \ce{Pb} in situ, so as to be able to accurately map sites for contamination. This study aims to address this need by examining the feasibility of developing an accessible and affordable Radio Frequency (RF) based wireless sensor that can monitor \ce{Pb}.


As a rapidly emerging technology, radio frequency-based wireless signals have shown capacity in material-level sensing in recent years, such as liquid classification/testing \cite{dhekne2018liquid,lobato2014wireless,ren2020liquid,dietz2002wireless}, food/fruit quality monitoring \cite{tan2018sensing,huang2011passive}. However, in the field of soil sensing, material-level sensing is still challenging due to the complexity of the soil medium \cite{xu2024survey}. 

In this study, we aim to examine the feasibility of measuring contamination of Pb in soil through RF-based technologies. The key idea of this study is that when RF signals, such as Wi-Fi (2-5GHz) and RFID (Radio Frequency Identification) System \cite{weinstein2005rfid} ($\sim$ 900MHz) are transmitted by an RF transmitter, they are affected by the medium through which these signals propagate \cite{seybold2005introduction}. The propagation medium reflects, refracts, or absorbs the RF signal, and this property varies depending on the composition of the medium. By examining the signal at the RF receiver, we can infer the composition of the medium. We have observed that different frequencies of RF signals react differently to different salts. Thus, a sensing system that incorporates RF signals at various frequency bands is proposed here. Our experiments include the following: First, software-defined radio is used to demonstrate that Pb (lead chloride salt) has a distinct effect on different frequencies' RF signals than \ce{NaCl} (sodium chloride, a common salt in soils) \cite{tavakkoli2010high}. Second, a commercially accessible RFID device is used to see if the aforementioned observations remain valid. Next, a simple regression system is developed to demonstrate the feasibility of measuring the contents of two different salts, in this case \ce{Pb(NO3)2} and \ce{NaCl} in the soil. We then built a machine learning model trained on 22 field samples and showed that we could classify \ce{Pb} level in soil (\revision{at the current U.S. EPA screening level of} 200ppm \revision{\cite{epa2024lead}}) with \revision{an accuracy of 72\% and recall} of 80\%. Finally, we discussed potential future work to make the sensor device robust and applicable to real-world soil conditions.

The main contributions of the work are as follows.
\begin{itemize}
    \item We introduce SoilScanner, the first RF-based soil components analysis tool that can detect \ce{Pb} contamination in soil. Existing RF-based wireless sensing systems have only been used for moisture and salinity sensing.
    \item We mathematically proved the feasibility of sensing different salts using RF signals. To our knowledge, this is the first work on modeling individual salt components in the soil.
    \item We open-source the first RF soil dataset that contains 23 lab-prepared ``control'' soil and 22 field soil samples. The dataset includes soil properties and RF signatures from our SoilScanner system.
\end{itemize}

\section{Related Work}


\textbf{Existing soil heavy metal (including Pb) detection: } Widely adopted high-accuracy soil contamination detection techniques are all non-RF based, including inductively coupled plasma mass spectrometry (ICP-MS) \cite{matong2016fractionation}, Atomic Absorption, X-Ray Fluorescence \cite{dos2000quantitative}, etc.  The trade-off of these high-accuracy methods is the high labor costs and lab test requirements. While the portable XRF instrument \cite{shefsky1997comparing} enables the in-situ testing capacity, its price is in the range of \$20,000-60,000, which is not accessible to most economically disadvantaged communities or gardeners.

Terahertz spectroscopy is another emerging technology utilizing signals between RF range and laser optical region for heavy metal detection in soil \cite{lu2022detection}. By analyzing the reflectance of terahertz radiation, researchers can generalize the pattern of the signal.  However, the terahertz sensor is also expensive (\$1,500-3,000), and the portability is also a concern.

\textbf{RF-based soil sensing: } We summarized the existing RF-based soil sensing techniques in Table \ref{tab:relatedwork}. Bechet et al. utilized a commercial off-the-shelf (COTS) RF system to test soil moisture \cite{aroca2016application}. Their work demonstrated the use of an RFID tag with a reader to estimate soil moisture. The moisture level is estimated from the signal power reflected from the tag. RFID has also been applied to greenhouse settings with improved accuracy \cite{wang2020soil}. However, salts in soil can affect the accuracy of single-band power-based moisture measurements because salt changes the electronic conductivity of the soil and affects the power of the signal received by the receiver. To overcome the salt interference of the power-based RF systems, researchers proposed methods that used the signal Time of Flight (ToF) instead of the power. Strobe \citep{ding2019towards} utilized a limited bandwidth WiFi frequency signal to measure the relative time of flight between the three antennas buried underground and then measured moisture by ToF and salinity by signal power. \revision{Note that the} salinity measurements reveal the total concentration of all salts in the soil, \revision{instead of } individual salt or contaminants. \revision{Additionally, Comet \cite{khan2022estimating} proposed a reflection-based RF sensing approach for detecting moisture without disturbing the soil, such as excavation or embedding antennas underground.}

To our knowledge, this study is the first attempt to differentiate individual salts in soil using RF signals, \revision{and this is an essential step for building an in-situ Pb measuring system in the future} 

\section{Background}

\subsection{RF signal propagation}

RF signals are electromagnetic waves in the frequency range of 3KHz to 300GHz, including the bands that carry data through wireless communication systems such as Wi-Fi, FM radio, and Bluetooth. An RF system consists of a sender that transmits an RF signal, a receiver that receives both the direct and indirect reflections of the RF signal and a transmitting medium in between. Consider an RF signal\revision{:} a single-tone cosine wave of amplitude $A_s$, frequency $f$, and phase $\phi$ transmitted by the sender, the signal received is given by the following equation:
\begin{equation} \label{eq_prop}
S_{r}(f, d) = A_{s}e^{-(\alpha + j\beta)d}
\end{equation}

Where $S_r$ represents the received signal, $d$ represents the distance traversed by the signal, $\alpha$ represents the attenuation coefficient (signal loss per unit of distance traveled caused by the medium), and $\beta$ represents the phase coefficient (phase shift per unit of distance traveled caused by the medium).

In any transmitting medium, the power of the signal decreases exponentially as distance ($d$) increases. If the transmission medium is solely air, the signal will spread out naturally and be attenuated along the free space path. For other transmission mediums, a portion of the signal is absorbed, and the rest either penetrates through or is reflected and scattered by the medium. The absorption/reflection ratio depends on the frequency of the transmitted signal and the characteristics of the medium. For example, a metal object is a good reflector of RF signal; RF signals of the lower frequency range can pass through an object while high-frequency RF signals are more reflected by materials.

Hence, the RF signal received by the receiver depends on three main factors: 1. frequency of the RF signal sent by the transmitter; 2. distance ($d$) between transmitter and receiver; and 3. characteristics of the transmission medium. In an RF sensing system transmitting signal and distance ($d$) between the sender and receiver are usually known, thus signal propagation will only be affected by the attenuation coefficient ($\alpha$) and phase coefficient ($\beta$) of the transmitting medium. This can be used to infer and differentiate the composition of different mediums. 

\subsection{RF-Signal propagation in soil}

In the context of soil medium, both moisture and salinity exert notable influences on the signal. Moisture affects the signal by predominantly changing the \textbf{dielectric constant $\epsilon'_{r}$} of the medium. The dielectric constant is a fundamental property that governs the capacity of a material to store electrical energy. The dielectric constant of pure water is relatively high, at 80, in contrast to other prominent constituents found in soil, such as sand (ranging from 3 to 5) and air (1). The significant disparity in the dielectric constant of water and other materials allows for the possibility of determining moisture level through dielectric constant measurement. Directly measuring the dielectric constant is difficult; however, it can be estimated by utilizing the apparent dielectric permittivity of the soil, represented as $\epsilon_a$, which can be determined through in-situ measurements. Apparent dielectric permittivity can be directly calculated through the velocity of the signal penetrating through the soil by equation $\epsilon_a = (c/v)^2$, where $c$ represents the speed of light. In practice, determining velocity $(v)$ involves measuring a signal's $ToF$ as it traverses through the soil. Given that the distance, denoted as $d$, is known, the velocity $(v)$ can be calculated by dividing the $d$ by $ToF$. After apparent dielectric permittivity is measured, moisture content can be determined using the Topp equation \cite{topp1980electromagnetic}. 

Soils commonly consist of both moisture and various substances, including salts like \ce{NaCl} and \ce{Pb(NO3)2}. The presence of salt in soil has an impact on its electrical conductivity \revision{($\sigma$)}, which refers to the soil's capacity to conduct an electrical current. Electrical conductivity impacts the signal power. In the controlled experiment, we observed that different frequency signals exhibit different responses to different salts. Therefore, a multi-band RF-based system is designed to understand the potential for differentiating various soluble salts in the soil, including Pb which is a ubiquitous contaminant.
\section{Design}

\begin{figure*}
    \centering
    \includegraphics[width=0.77\linewidth]{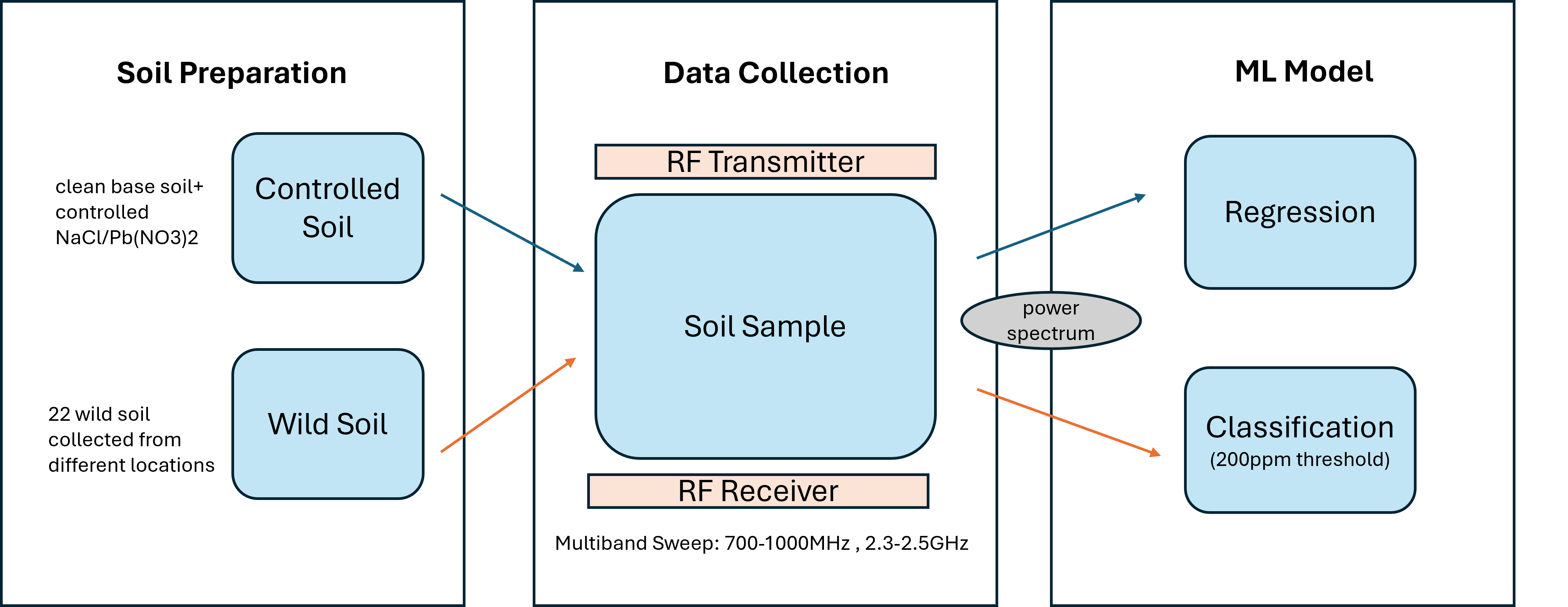}
    \caption{Flow chart of SoilScanner.}
    \label{fig: flow}
\end{figure*}

An overview of the SoilScanner system is shown in  Fig. \ref{fig: flow}. The prepared soil (detailed in the implementation section) is placed in a 1L plastic container between the fixed transmission and receiving antenna. The transmitter emits short periods of single-tone RF signal at 700-1000MHz and 2.3-2.5GHz with a step of 0.5MHz. Then, the received power spectrum is passed to the machine learning (ML) model for further analysis.

\subsection{RF signal affected by different soil salts}

The insight of the multi-frequency-based power spectrum collection system can be summarized as follows.

\begin{enumerate}
  \item \textbf{The electrical conductivity ($\sigma$) \cite{lewis1980practical} of each salt exhibits variations based on its density and exhibits distinct responses to alterations in the frequency of the surrounding electromagnetic field.}

  The relationship between the density and electrical conductivity of different salt solutions exhibits variations when subjected to constant magnetic field frequency, temperature, and pressure.  The varying current-carrying capacities and saturation levels of different ions offer an explanation. 

  \item \textbf{The impact of the ambient magnetic field on various salt solutions varies.} For example, the application of a magnetic field leads to an increase in the conductivity of \ce{NaCl}, \ce{MgCl26H2O}, and \ce{KCl}. The solubility of \ce{MgSO4}, \ce{CaCl22H2O}, and \ce{NaHCO3} exhibits a contrasting decrease. \cite{guo2016frequency}

\end{enumerate}

In a controlled environment with constant temperature and pressure, the conductivity of each salt $i$ can be expressed as a function $\sigma_i(f)$ of the varying frequencies $f$ of the applied electromagnetic wave. In addition, it is important to note that soil is composed of a diverse array of salts. The electrical conductivity ($\sigma$) of soil is a comprehensive measure that encompasses the collective impact of all these salts.
$$\sigma_{soil}(f) = \sigma_{S{(\vec{d})}}(f)$$ 
The density of each salt is denoted by the vector $\vec{d}$. And the function $S$ is the combination function of all salts' conductivity.

When a soil sample is placed between a transmitting and a receiving antenna, the signal attenuation percentage can be measured and defined by the attenuation factor. \cite{ding2019towards}

\begin{equation} \label{eq_alpha}
\alpha = \frac{2{\pi}f}{c} \sqrt{\frac {\epsilon'_{r}}{2}[\sqrt{1+\tan^2{\delta}}-1]}
\end{equation}

where $\epsilon'_r$ is the dielectric constant, $c$ is the speed of light, and $\tan\delta$ is the loss tangent of the transmitting medium soil. 

\begin{equation} \label{eq_loss tangent}
\tan{\delta} = \frac{\epsilon''_{r}(f)+\frac{\sigma_{soil}(f)}{2{\pi}f\epsilon'_{0}}}{\epsilon'_{r}} = l
\end{equation}

where $\epsilon''_r(f)$ is the dielectric loss, which is the energy required to heat a dielectric material in an alternating electric field. Loss tangent is noted using parameter $l$.

Therefore, at each frequency, when dielectric constant $\epsilon'_{r}$ is known, the power of the signal transmission is related to the conductivity of the materials and the dielectric loss of the materials.

$$Power_{soil}(f) \propto l_{soil}(f) = l_{S{(\vec{d})}}(f)$$

For each salt $i$, the loss tangent varies as the frequency changes due to the fact that the dielectric loss of each salt varies as the frequency changes \cite{levitskaya2019parameters} as well as insight 1 and 2 that electric conductivity of each salt varies differently when at different frequency magnetic field.

Since the objective is to determine each salt concentration in the soil, the question can be expressed as the following mathematical problem. Outputs consist of the densities $\{d_i\}_{all\ salts}$ of each distinct salt $i$. The input is the power transmitted and received at each frequency $f$, $Power_{soil}(f)$, which is proportional to the loss tangent of the combined salts when the moisture level and environmental factors (temperature, pressure) are known.

A learning system can then be developed. The system will accept $Power_{soil}(f)$ at distinct frequencies $f$ as inputs and will output the density of each salt in the soil. The number of inputs will vastly exceed the number of outputs, resulting in a high capacity for estimation.





\subsection{Learning algorithm for lab prepared controlled soil samples}

\revision{We begin our analysis with controlled soil samples prepared in the lab. For these samples, we spiked "clean soil" with two different salts—\ce{NaCl} and \ce{Pb(NO3)2}, in varying quantities. Details of the soil preparation are provided in Section \ref{ssec:soil_preparation}}.


\revision{We first analyze the power spectrum of the data and then apply two key measures to build a regression model that estimates the concentration of Pb in the soil.}

\revision{This section} aims to develop a method for choosing parameters to distinguish between two different salts using data analysis. \revision{Since all other properties of the lab-prepared soil were the same except for the \ce{NaCl} or \ce{Pb(NO3)2} content, a narrow frequency band is sufficient to characterize the variation in Pb concentration. The selected frequency band showed the most significant correlation with Pb concentration.} 

\revision{Specifically,} we introduce two key measures: Diff800 and Diff2300. Diff800 shows how much the power of the signal changes between 810.1MHz and 790.1MHz, while Diff2300 shows the same between 2408.6MHz and 2401.1MHz. \revision{By subtracting readings between two adjacent frequencies, the changes in the slope caused by Pb become more apparent and can be better evaluated. This process emphasizes the relative relationship between frequencies.
}

By concentrating on these measures, we hope to go beyond environmental influences and better understand how soil composition affects RF signal behavior. This structured approach provides a clear method for selecting parameters, making it easier to classify and differentiate between different salt types in soil samples. This contributes to the advancement of RF signal analysis in environmental research and monitoring.

\begin{figure*}
    \centering
    \begin{subfigure}{0.40\linewidth}
        \centering
        \includegraphics[width=\linewidth]{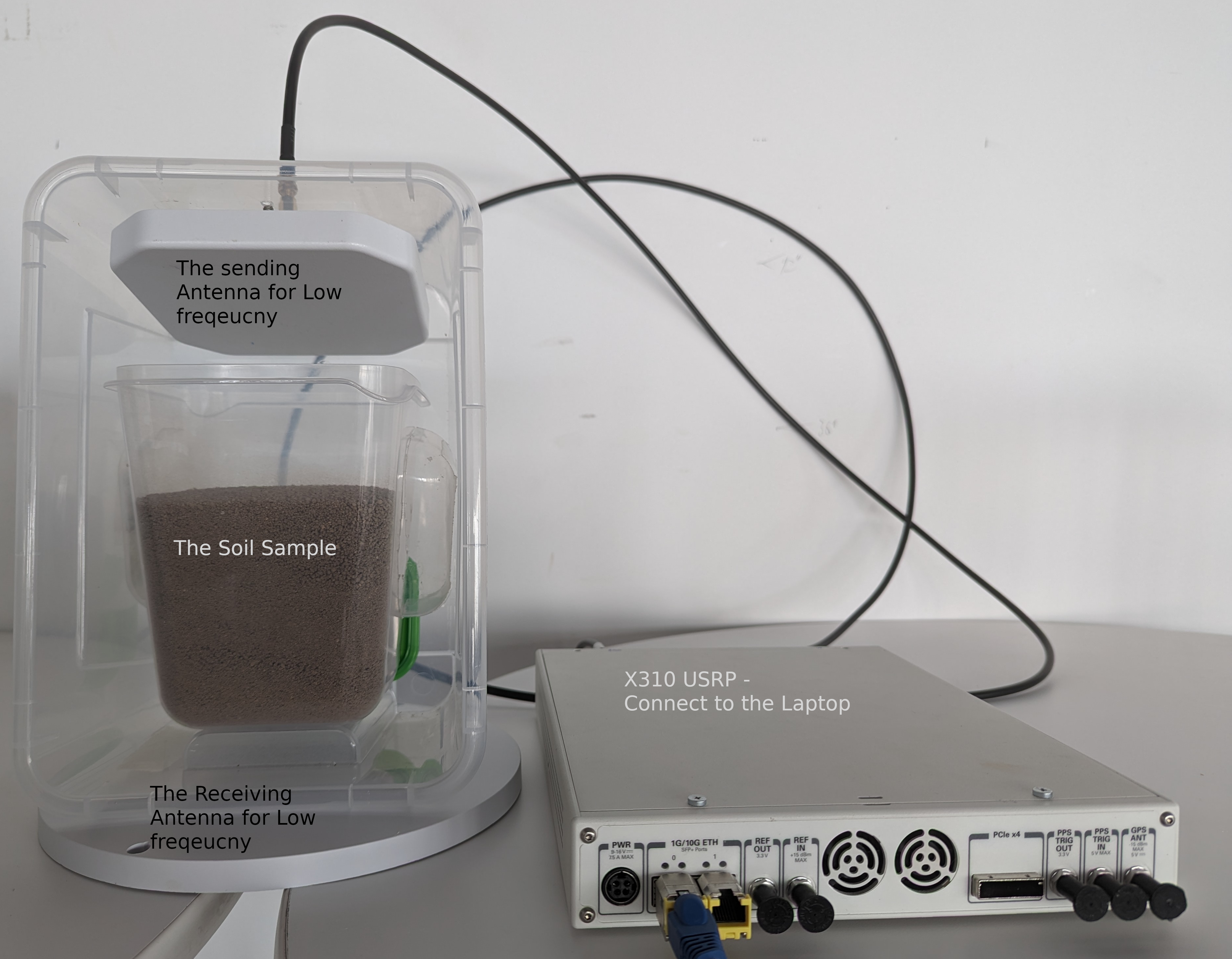}
        \caption{USRP low-frequency range setup.}
    \end{subfigure}
    \begin{subfigure}{0.4137\linewidth}
        \centering
        \includegraphics[width=\linewidth]{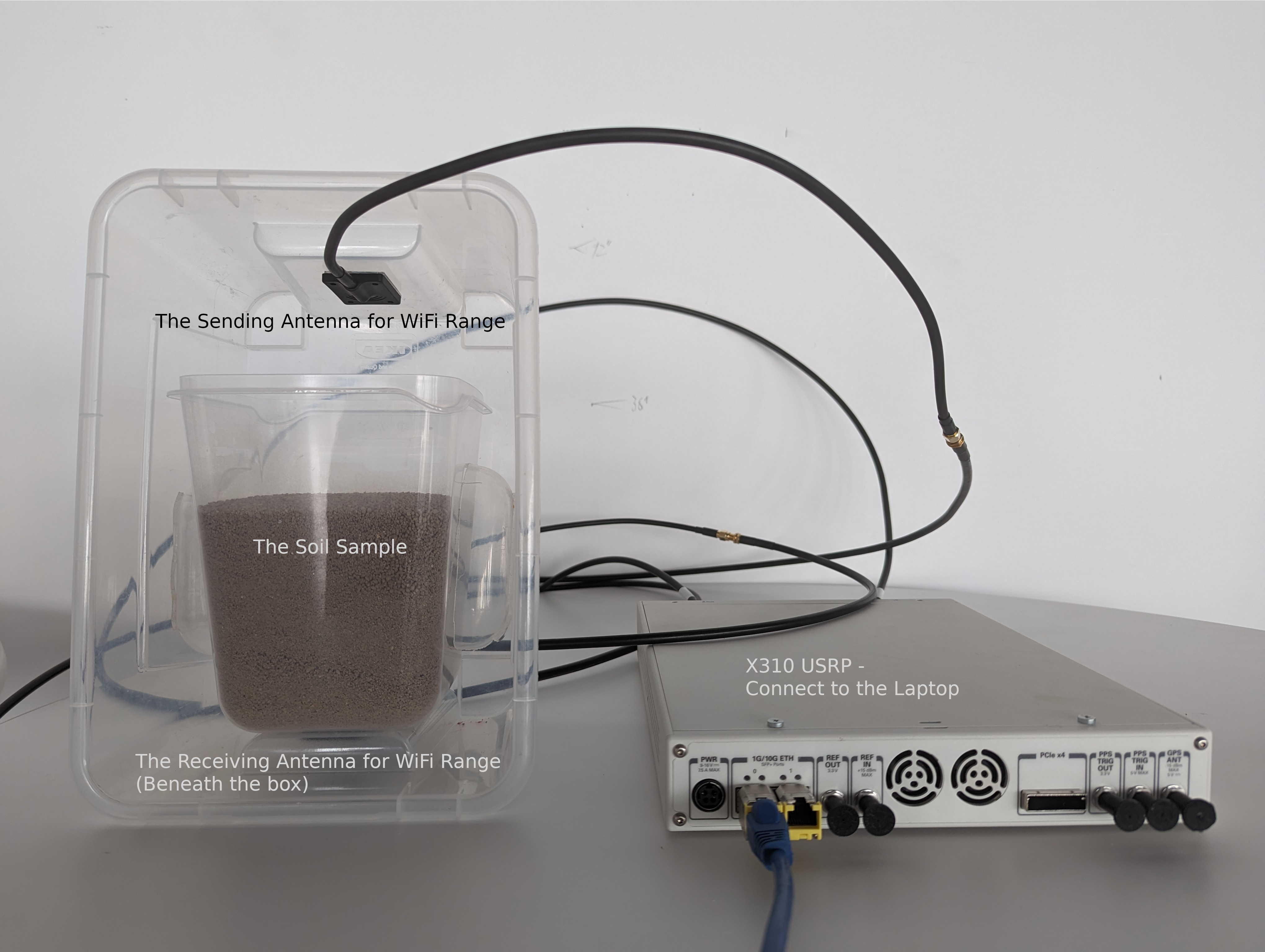}
        \caption{USRP high-frequency range setup.}
    \end{subfigure}
    \caption{USRP setup.}
    
    \label{fig:usrp_setup}
\end{figure*}

\subsection{Learning algorithm for uncontrolled field samples}

\revision{Next, we analyze 22 field soil samples collected from various locations around a large metropolitan area. The details of the sample preparation are provided in Section \ref{ssec:soil_preparation}}. An ML-based model, shown in Fig. \ref{fig: Model_wild}, was built for binary classifying soil by criteria of 200ppm of Pb, the new screening level for Pb in the residential area set by U.S. EPA in early 2024 \cite{epa2024lead}.

\subsubsection{Feature Engineering}
 To make the learning process easier and reduce near-frequency data redundancy, we design the following feature engineering techniques:
\begin{itemize}
    \item \textbf{Frequency hopping}: instead of using all the frequencies', we select one frequency every certain interval.
    \item \textbf{Frequency aggregation}: similar to the hopping technique, we select the average value in that interval instead of selecting only one frequency every certain interval.
    \item \textbf{Weighted frequency aggregation}: as the name suggests, instead of giving every interval a uniform weight, we selectively weigh the intervals where we can see the most variance in the data.
\end{itemize}

\subsubsection{Train-test-split}

Before splitting the dataset, we binary encoded each sample's \ce{Pb} level as 0 and 1 by separating at the 200 ppm threshold. After encoding, 10 samples were classified as exceeding the amount of \ce{Pb} (1), and 12 samples were classified as within the safe threshold (0).

We utilized Leave-one-out cross-validation (LOOCV) for the train-test-split for our dataset. LOOCV is a technique for evaluating a machine learning model's performance by training it on all but one data point and then testing it on the one held-out data point. This process is repeated for each data point, and the average performance across all iterations is used to estimate the model's generalization error. For smaller datasets, LOOCV has the benefit of providing an unbiased estimate of the model's performance, and it maximizes the information used for model assessment without sacrificing data. In our system, we repeat LOOCV $k$ times and calculate the average performance of all $k$ iterations for reduced bias of the noise added in the Data Augmentation step below.

\subsubsection{Data Augmentation for Training Data}
For the model to be location-invariant and robust against noise, we need to increase the number of training samples. To do that, we designed a simple data augmentation technique that can up-sample the training dataset by a factor of $r$. First, all training samples were copied $r$ times. Then, random noise, $n$, was added at each frequency power of each sample where, $n \sim \mathcal{N}(0,\,\sigma^{2})$, and $\sigma \sim \mathcal{U}(0,\,3)$. The reason for such a range for the standard deviation is that when we tested the same sample at different locations, we observed a standard deviation in the data that is similar to that range. $$X \xrightarrow[]{\text{up-sample by r}} X_u \xrightarrow[]{\text{add Gaussian noise}} X_{augmented}$$

\subsubsection{Model Training}

To achieve high robustness and performance, we utilized a soft voting classifier. A voting classifier is a type of ensemble learning method in which several base models are trained independently on the same dataset, and their predictions are combined through a voting mechanism to make the final prediction. In soft voting, the output probabilities of each base model are averaged for each class, and the class with the highest average probability is chosen as the final prediction. This method takes into account the confidence of each base model in its prediction rather than just the raw class labels. By combining multiple models with potentially different strengths and weaknesses, a voting classifier achieves better generalization and robustness compared to any individual base model. Our diversified base model includes Logistic Regression, SVM, Decision Tree, and Naive Bayes. Detailed model implementation is \revision{shown in Fig. \ref{fig: Model_wild}.}


\section{Implementation}

\subsection{Hardware} 
A software-defined \revision{Universal Software Radio Peripheral} (USRP), USRPX310 with two UBX-160 daughterboards operating at wideband frequencies ranging from 10MHz-6GHz was used. One daughterboard is the sender and is connected to a Laird S9025PL antenna for testing frequencies less than 1GHz. For frequencies above 1GHz, the sender daughterboard is connected to Taoglas Limited’s ultrawideband FR4 antenna. The other daughterboard is the receiver and is connected to an EC742IOM06H-PIM-NF wideband antenna for frequencies less than 1GHz and Taoglas Limited’s ultra-wideband FR4 antenna for testing frequencies above 1GHz. The two antennas are separated by 18cm and attached to the inner top of a plastic box with a dimension of 15cm*15cm*20cm. During an experiment, the sender antenna is attached to the top of a box, and the receivers are attached to the bottom. For all experiments, the soil is placed in a closed plastic box in the middle, 5-8cm from the sender and 2cm to the receiver, as shown in Fig. \ref{fig:usrp_setup}. The USRPX310 device is connected to a Lenovo Thinkpad laptop with a 1 GHz ethernet interface. 

\subsection{Software} 
The data acquisition system is programmed using GNU Radio version 3.10.5.1, an open-source software that controls the USRP with a graphic interface. 

A sending signal comprises a base signal modulated on a carrier wave. The base wave is a cosine wave at a frequency of 100k. For each frequency range (modulated wave frequency 700-1000Mhz and 2300-2500Mhz), the initial carrier wave frequency is set to be the lower bound of that specific range, with a sample rate of 1 million samples per second and a gain value as the maximum value of the antenna capacity, 6dBm. The frequency shift is done on the carrier wave controlled by a global parameter $freq\_contorl$, updated at the receiver side.

On the receiver end, the raw signal is first received and demodulated at the USRP source block. Then, it is passed on to a Fast Fourier Transform (FFT) unit, which performs an FFT on every 1024 samples received. The FFT converts the time domain signal into the frequency domain. The power and standard deviation are then calculated by averaging 100 data points ($\approx 0.2$sec) processed from the FFT processing unit. If the standard deviation exceeds a threshold of $0.02$dBm, the program will recollect the signal for that specific frequency. The system will skip to the next frequency if it misses the standard threshold five consecutive times and report to the user. If successful, the data collection program will increase the $freq\_control$ variable by a step of $0.5$MHz and collect data for the next frequency. 

After all the data on the frequencies are collected, both R and Python are used for data analysis.

\subsection{Soil Preparation} \label{ssec:soil_preparation}
\revision{For evaluating our system, we prepared two sets of samples: 23 lab-prepared controlled soil samples, and 22 field-collected uncontrolled soil samples.}

\textbf{Lab prepared controlled soil samples: } For the soil experiments, we chose loamy sand topsoil with low salt (<75 ppm), low \ce{Pb} content (<5 ppm), and low organic content (<3\%) as the base soil. The soil texture (loamy sand) is a common natural soil in the New York City \revision{(NYC)} metropolitan area. The soil was air-dried for five days, passed through a 2mm sieve, and fully homogenized before being used. This soil was then divided into 1kg each and placed in a 1.7L IKEA Pruta plastic food container. The Pruta containers are built with thin plastic walls, reducing the signal loss when transmitting through the container. We then prepared two different salts.  \ce{NaCl}  (table salt, with $97-99\%$ \ce{NaCl}) and \ce{Pb(NO3)2} (a common soluble \ce{Pb} compound). It should be noted that once \ce{Pb(NO3)2} is mixed with soil, Pb can react with various components in the soil and become less soluble.

Finally, we carefully spiked each sample using varying amounts of the above two salts, separately or as a combination of two salts. The different compositions are listed in the Table \ref{table:tab1}. The soil in each container was thoroughly mixed after spiking. All these samples were rested for at least 24 hours before testing.

\begin{table}[ht]  

\caption{Soil variables for experiments in this study.} 
\centering 
\resizebox{0.5\textwidth}{!} {
\begin{tabular}{ c c c c c r r r r r } 
\hline\hline   
 Sample ID & Base Soil mass(g) & \ce{NaCl} (ppm) & \ce{Pb(NO3)2}(ppm)
\\ [0.5ex]  
\hline   
1-7 & 1600  & [0,50,100,200,400,1000,2000] & 0 \\
8-14 & 1600 & 0 & [200,400,1000,2000] \\
15-23 & 1600  & [100,400,2000] & [100,400,2000] \\
\hline 
\end{tabular} 
}

\label{table:tab1}
\end{table} 

\textbf{Field collected uncontrolled soil samples: } 22 distinct soil samples were collected from the field (gardens, farms, and other land uses in the NYC metropolitan area). \revision{The composition of these soil samples, including moisture, organic content, gravel, salt, pH, etc., was carefully examined at the Urban Soils Lab at Brooklyn College. The uncontrolled samples offer substantial diversity beyond the lab-spiked samples. These analyses revealed significant variability among the samples, which will be discussed in detail in Section \ref{ssec:composition}.}


\section{Evaluation}
We first demonstrate the robustness of the sensing system through its reprehensibility through two tests: the remounting test and the location test. Then, we analyze the effects of \ce{Pb(NO3)2}, \ce{NaCl}, and their combined effect on signals using controlled soil samples. Next, we verify whether these patterns are also observable in COTS RFID devices. Subsequently, we evaluate the performance of regression models built on controlled soil samples. Finally, we evaluate the machine learning model's performance for the uncontrolled field soil samples. 

\begin{figure}[t]
    \centering
    \includegraphics[width=0.9\linewidth,height=0.5\linewidth]{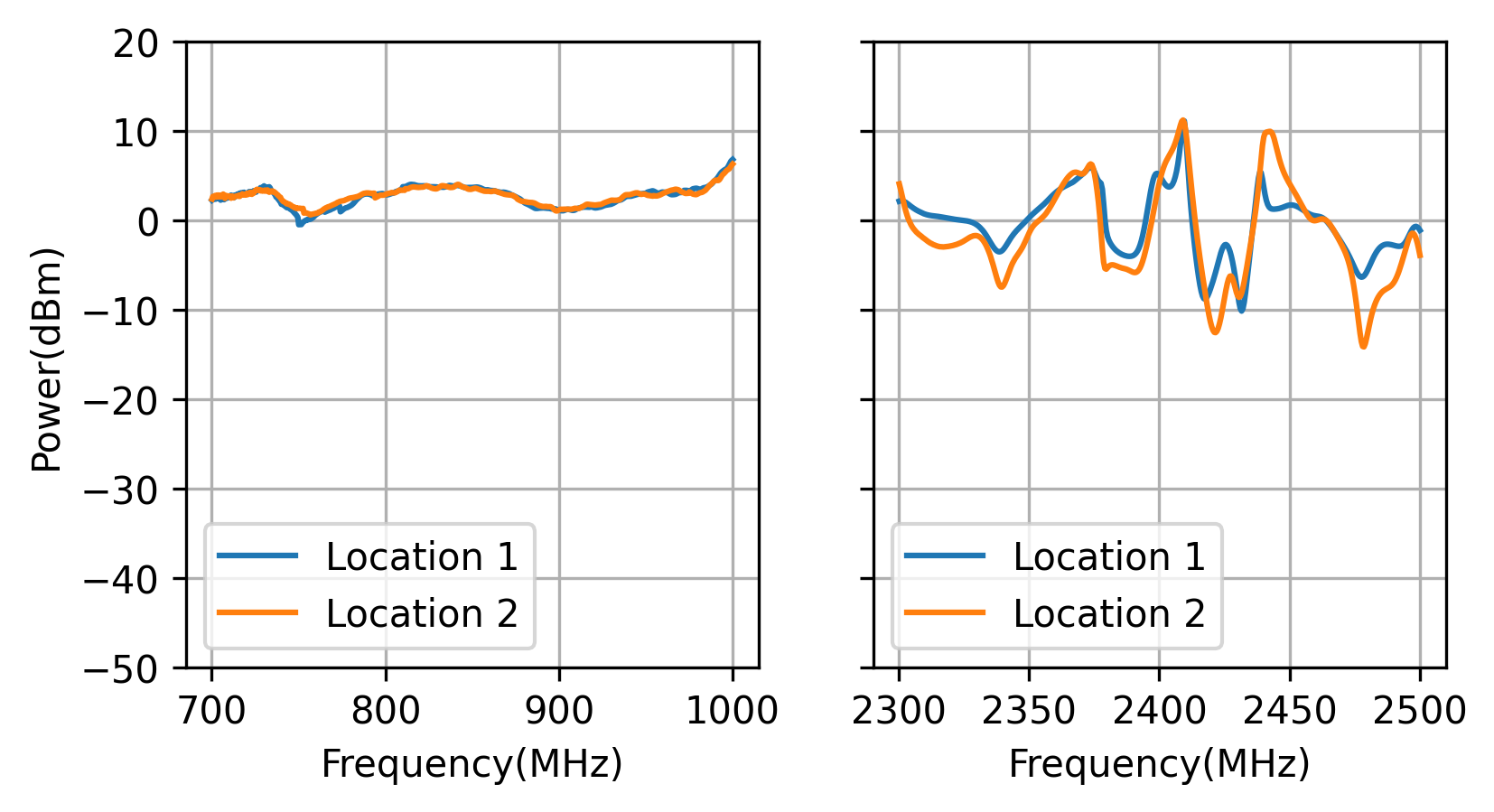}
    \caption{Location dependency test.}
    \label{fig: loc_dep_test}
\end{figure}

\subsection{Location Robustness}
One field sample's data was collected at two locations and then compared to understand the effect of location dependency in our RF sensing system. After removing the background signal, they are plotted in Fig. \ref{fig: loc_dep_test}. Except for the small variation towards the higher frequencies, both signals are well-matched. The variation is expected because of interference with the WiFi signals present in the environment. However, this higher frequency interference can be further reduced by placing the SoilScanner system inside a metal box. 

\begin{figure}[t]
    \centering
    \includegraphics[width=0.9\linewidth,height=0.5\linewidth]{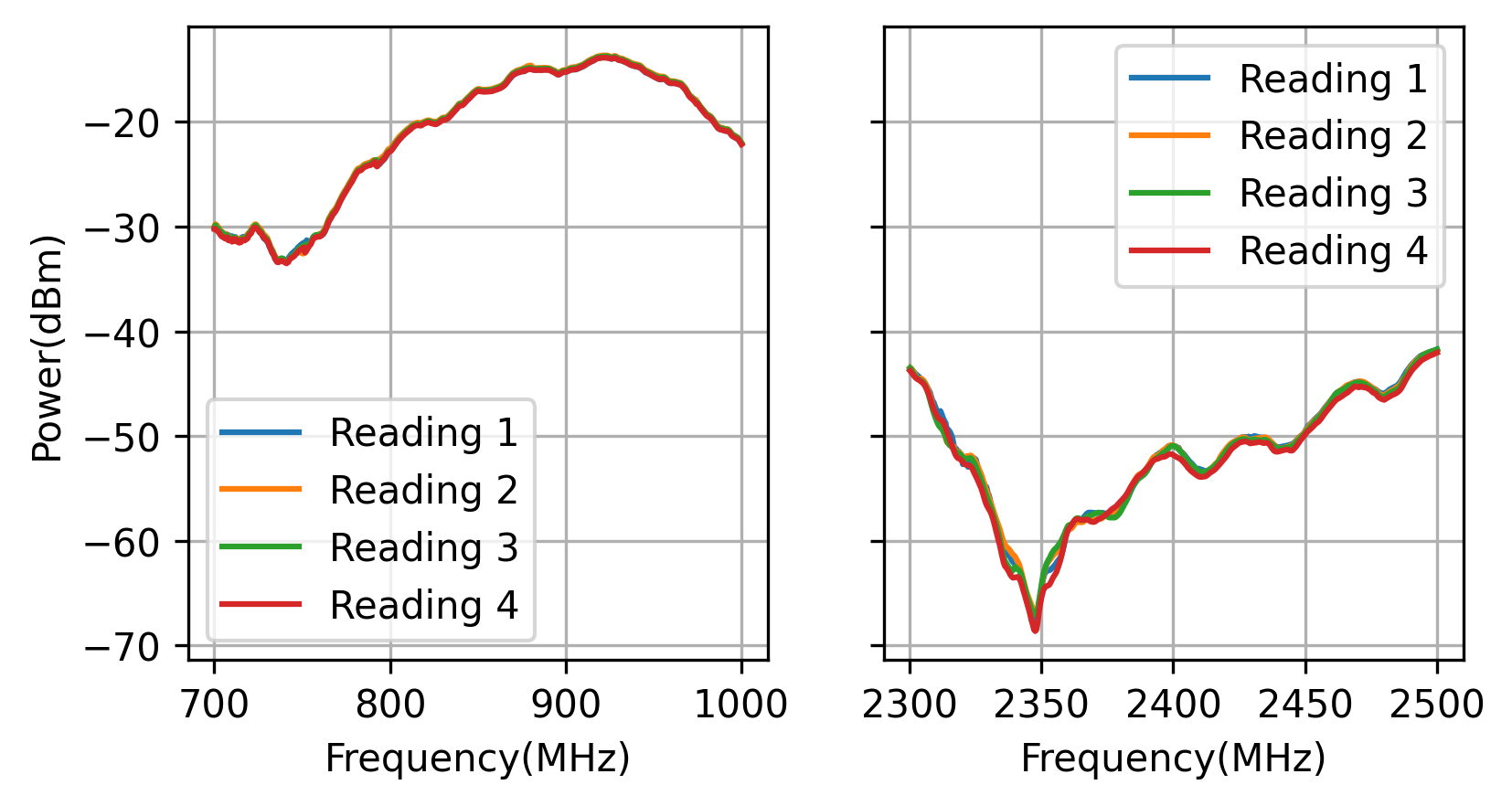}
    \caption{Remounting test.}
    \label{fig: remounting_test}
\end{figure}

\begin{figure*}
    \centering
    \includegraphics[width=\linewidth]{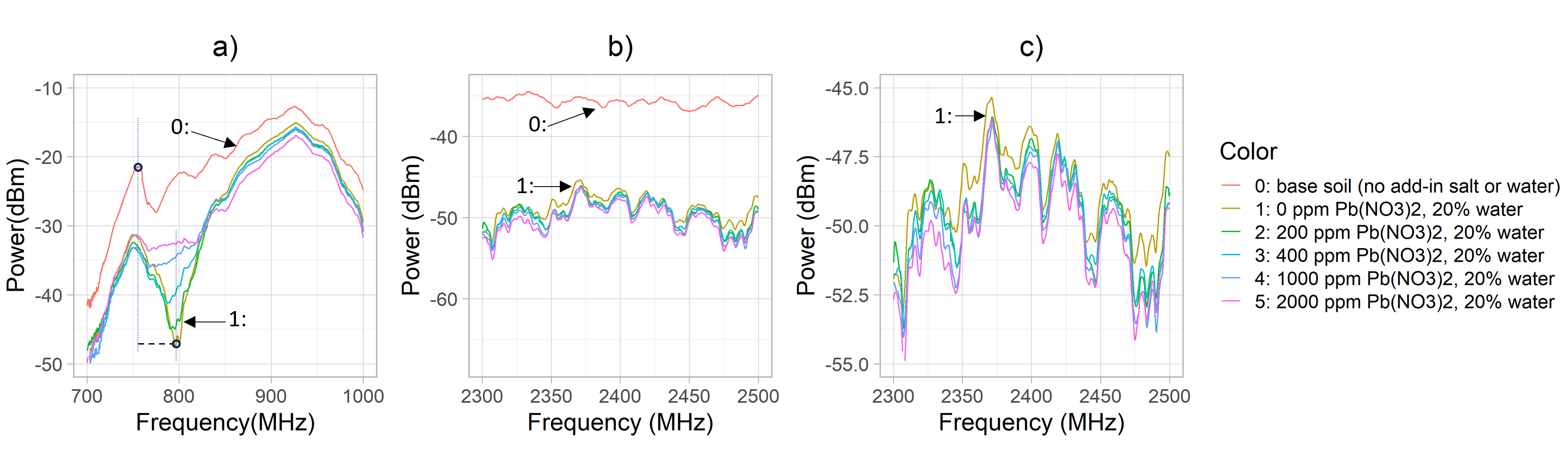}
    \caption{Received signal power (in dBm) variation with testing frequency a) 700-1000MHz, b)2300-2500MHz. Different colors represent different samples with varying \ce{Pb(No3)2} contents. c) is a zoomed-in view of b).}
    \label{fig: USRP_Lead}
\end{figure*}

\begin{figure*}
    \includegraphics[width=\linewidth]{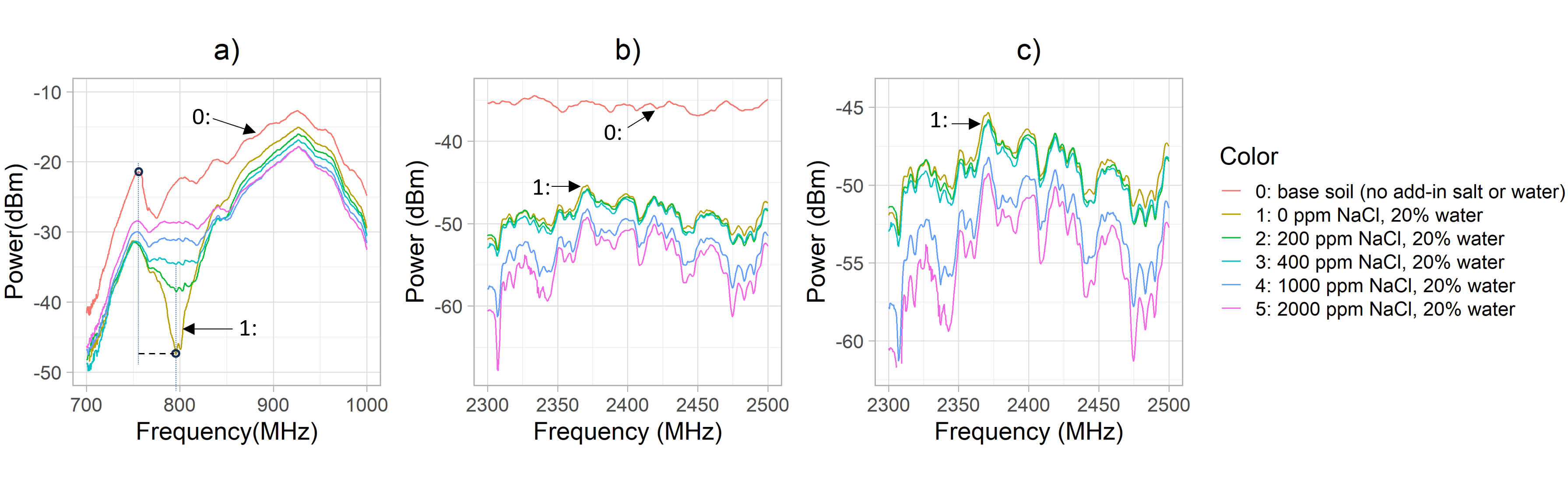}
    \caption{Received signal power (in dBm) variation with testing frequency a) 700-1000MHz, b)2300-2500MHz. Different color represent different samples with varying \ce{NaCl} contents. c) is a zoomed-in view of b).}
    \label{fig: USRP_NaCl}
\end{figure*}

\subsection{Remounting}
A remounting test was performed to check the performance of our system. One field soil sample was tested four times at the same place, remounting the box every time between data collection. The signals are plotted in Fig. \ref{fig: remounting_test}. As evident from the data, there is no visible difference in the received signal in the four different trials.


\begin{figure}
    \includegraphics[width=\linewidth]{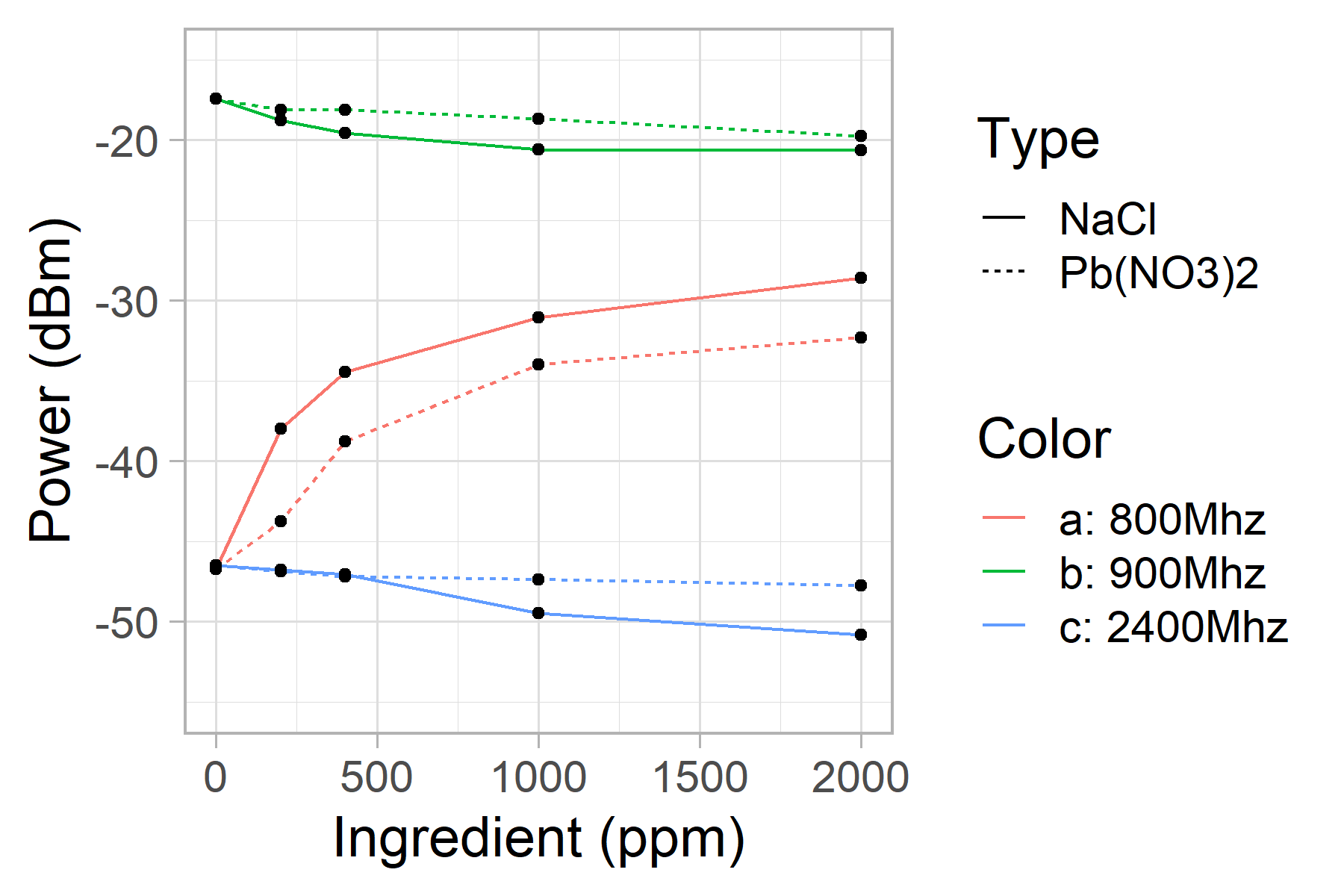}
    \caption{Received signal power (in dBm) variation with \ce{NaCl} and \ce{Pb(NO3)2} concentrations. Different colors represent different frequencies of the tested signal.}
    \label{fig: USRP_NaCl_PF}
\end{figure}

\begin{figure*}[t]
    \includegraphics[width=\linewidth]{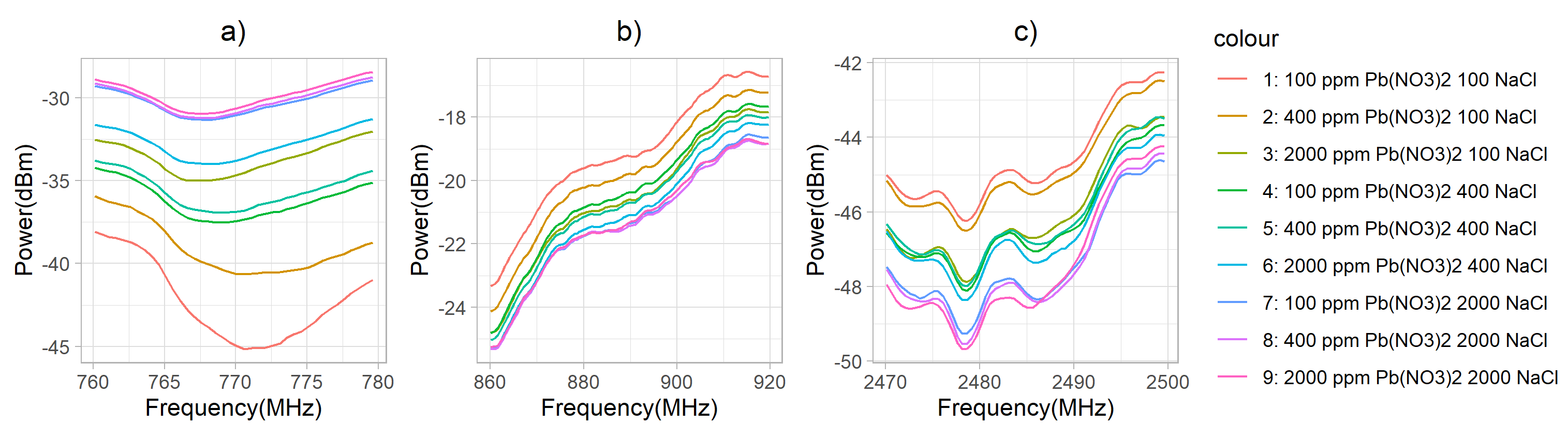}
    \caption{Different frequency range (x-axis) versus the RSSI (y-axis).}
    \label{fig: USRP_Mixed}
\end{figure*}

\subsection{Analysis on lab prepared controlled soil samples}

\subsubsection{RFID and \revision{WiFi} range signals for soils spiked with \ce{Pb(NO3)2}}

Soil samples spiked with only \ce{Pb(NO3)2} (samples 15-18, see Table \ref{table:tab1}) were first tested with the RF setup at two distinct frequency ranges: 700-1000MHz with the Laird antenna and 2.3-2.5GHz with the WiFi high-frequency antenna. In addition to the spiked samples, two soil samples were tested to establish a baseline - 0: the base soil sample with no added water represented by the red line, and 1: the base soil sample with $20\%$ of water represented by the lime green line. The average power of the received signal is plotted against frequency and shown in Fig. \ref{fig: USRP_Lead} and the average power of the received signal is plotted against \ce{Pb(NO3)2} concentration for selected frequency bands in Fig. \ref{fig: USRP_NaCl_PF}. These results, while obtained in the same setup, were repeated over multiple days and were found to be consistent. 

 In Fig. \ref{fig: USRP_Lead}a, When comparing the curve for the base soil (no water) and the curve for all other soils with water (moisture > 0), peaks for the latter slightly shifted towards 800MHz. This is because moisture can lead to higher permittivity of soil and a decrease in the speed of the signal that causes the peak shift. This is consistent with the findings in the literature that measured moisture content using RF technologies \cite{ding2019towards}. 
 In Fig. \ref{fig: USRP_Lead}a, if focused on 800MHz, The power of the received RF signal increases with \ce{Pb(NO3)2} concentration in soil (Fig. \ref{fig: USRP_NaCl_PF}). The magnitude of increase between samples (e.g., $\sim$ 5dBm for \ce{Pb} concentration from 200ppm to 400ppm) is significant. This is because, with an increase of every 6dBm, the signal's amplitude is doubled. However, at frequencies higher than 825MHz, the received RF power appears to decrease with increased soil \ce{Pb} content, shown as line b in Fig. \ref{fig: USRP_NaCl_PF}. This shows that the effect of \ce{Pb(NO3)2} on received RF power varies at different frequencies, even within the same RFID band in the current test setup.

Fig. \ref{fig: USRP_Lead}b plots the average power of the received RF signal in the WiFi frequency range (2.3–2.5GHz). It should be noted that the received RF power of base soil is very high (more than 10dBm higher) compared to all the other samples with moisture. Fig. \ref{fig: USRP_Lead}c is a closer view of all the sample curves with 20\% moisture. It can be seen that the power of the RF signal decreases with the increase in \ce{Pb} content in the soil. This pattern is consistent within this frequency range. However, the magnitude of the decrease in signal power due to \ce{Pb(NO3)2} is slightly smaller than that in the RFID 900MHz band. Selected 2.4GHz band signal power response to soil \ce{Pb} concentrations is plotted as line c in Fig. \ref{fig: USRP_NaCl_PF}. Here the difference in reflected signal power between soil samples with 2000ppm \ce{Pb} and that with no \ce{Pb} is only about 3dBm.  

This experiment demonstrates that variations in salt concentration impact the power of the received RF signal, and it varies at different frequencies.

\subsubsection{RFID and WIFI range signals for soils spiked with \ce{NaCl} \label{subsection: nacl_analysis}}

The above experiment was repeated with soil samples spiked with only \ce{NaCl} salt, and the results are shown in Figs. \ref{fig: USRP_NaCl} and \ref{fig: USRP_NaCl_PF}. Overall, in the 700 to 1000MHz range, the observations are similar to \ce{Pb(NO3)2} spiked soils. For example, soil moisture shifted the peak towards 800MHz (Fig. \ref{fig: USRP_NaCl}a). The power of the RF signal increases with the \ce{NaCl} content in soil at around 800MHz but decreases when above 825MHz (Fig. \ref{fig: USRP_NaCl}a and Fig. \ref{fig: USRP_NaCl_PF}). 

Within the \revision{WiFi} frequency range (Figs. \ref{fig: USRP_NaCl}b and \ref{fig: USRP_NaCl}c),  received power decreases with increasing content of \ce{NaCl}, which is consistent with findings from the prior experiment with \ce{Pb(NO3)2}.  However, the magnitude of the decrease is higher for \ce{NaCl} than \ce{Pb(NO3)2}; \revision{t}his can be observed from the greater slope of the lines c in Fig. \ref{fig: USRP_NaCl_PF}.






\subsubsection{RFID and \revision{WiFi} signals affected by a combination of \ce{Pb(NO3)2} and \ce{NaCl}}

The experiments with individual \ce{Pb(NO3)2} and \ce{NaCl} showed that RF signals at different frequencies have different sensitivities to the two studied salts.

However, most soils from the field usually contain many different types of salt. So, an additional experiment was performed with soil samples spiked with varying concentrations of both \ce{NaCl} and \ce{Pb(NO3)2} (Table \ref{table:tab1}, Samples 21-29). 

Fig. \ref{fig: USRP_Mixed} shows the power for the RF signal for all frequencies. RFID frequencies are plotted in two bands: 760-780MHz and 860-920MHz. The power increases with salt concentration in the 700-800MHz band, while the opposite trend can be observed in the 800-900MHz range. In the 2.4GHz spectrum, the signals are grouped into three clear ranges where the power is more attenuated when the \ce{NaCl} content is at 2000 ppm. This is similar to the observation in the previous experiment (see \revision{S}ection \ref{subsection: nacl_analysis}) where \ce{NaCl} affects the RF signal more than \ce{Pb(NO3)2} at 2.4GHz. This experiment shows that the effect of individual salts on reflected power signals is maintained with similar trends even when other salt(s) are present in the soil. While the effect by \ce{Pb(NO3)2} was maintained at the lower frequency band (700-900MHz), the effect by \ce{NaCl} was maintained at the higher frequency band (2.4GHz).

\begin{figure}
    \centering
    \includegraphics[width=0.95\linewidth]{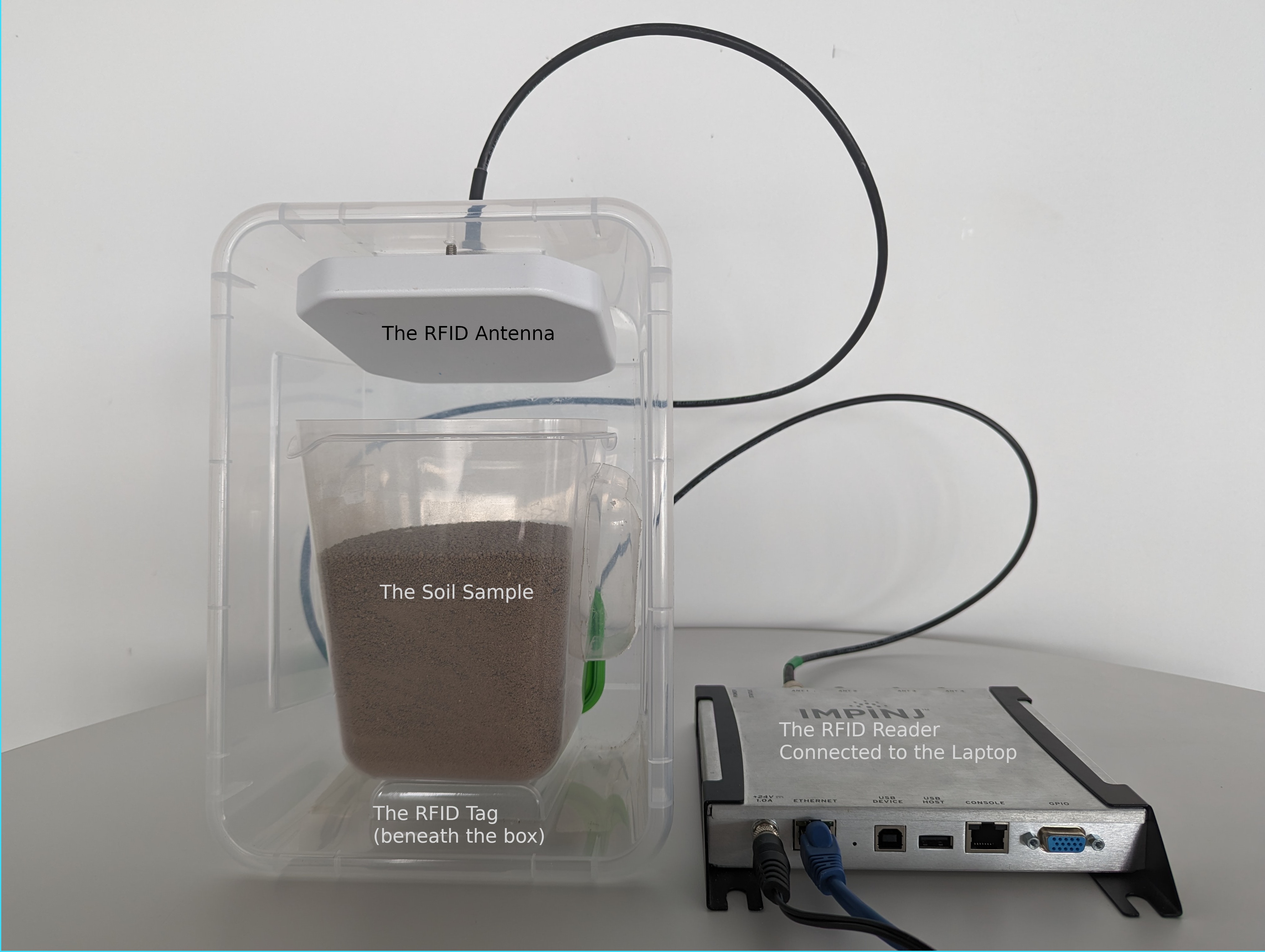}
    \caption{Experiment setup with commercial RFID.}
    \label{fig: Lead_possible}
\end{figure}

\subsubsection{Potential on commodity RFID}
The above experiments were performed on expensive software-defined radios. The same experiment with a commercial RFID setup was repeated to prove the feasibility of using commercial off-the-shelf devices.

\begin{figure}
    \includegraphics[width=\linewidth]{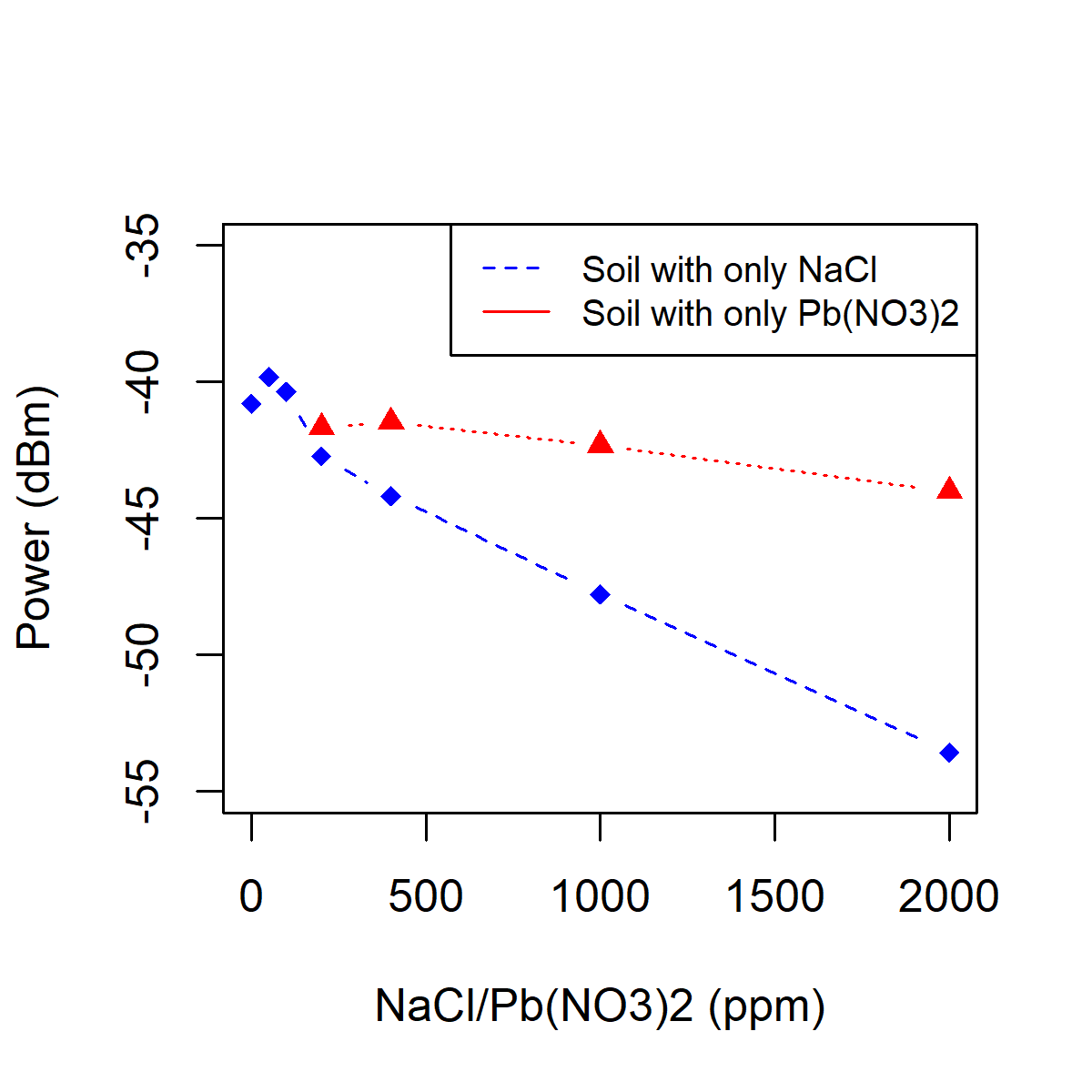}
    \caption{x-axis is the content of \ce{NaCl} or \ce{Pb(NO3)2} in ppm, and y-axis is the power reading \revision{of RFID setup} in dBm. The red line represents \ce{Pb(NO3)2} samples with $20\%$ moisture, and the blue line represents \ce{NaCl} samples with $20\%$ moisture.}
    \label{fig: RFID_salt}
\end{figure}

In this setup, a commercial Impinj R420 RFID tag Reader (Rev 3), which operates at 918\revision{-}926MHz, was used. No additional hardware-level modification for the RFID reader was needed. The Laird S9025PL \revision{a}ntenna was connected to the RFID reader using a Vulcan RFID 6 ft \revision{a}ntenna \revision{c}able and installed on the inner top of a plastic box with a dimension of 15cm*15cm*20cm (same as the USRP setting). The antenna has a 5.5dB gain and an elevation/azimuth \revision{b}eamwidth of 100 degrees. One EPC C1 GEN2 73X17mm UHF \revision{t}ag is attached to the inner bottom of the box, as shown in Fig. \ref{fig: Lead_possible}. The reader and the tag are separated by 18cm. We used Octane SDK with Java and ItemTest, an Impinj software, to control the reader activity. The reader's power was set at 30dB, and Rx sensitivity was set at -80dB. Each data collection session was 10 seconds, with tag-read activity occurring at least 100 times. One average power is calculated using all the power readings of the specific tag (attached to the box) within the 10s period.

Fig. \ref{fig: RFID_salt} shows the RF signal's power for different soil samples with commodity RFID setup. In a commercial RFID setup, only one average power value was recorded for each soil sample. For both \ce{NaCl} and \ce{Pb(NO3)2}, the power of the received signal is negatively proportional to the amount of salt in the sample. Furthermore, the slope representing how sensitive the RF power is changed due to the presence of salt is clearly different for the two studied salts. It is important to note that this observation is the same as that of the expensive software-defined radio setup at the 900MHz range.

\subsubsection{Regression on controlled soil samples}

The regression model shows the feasibility of separating \ce{Pb} inside the soil. The $R^2$ of each pair of features and ingredient is shown in Table \ref{table:tab2}.  Diff2300 can be used to estimate the level of \ce{NaCl} because the linear relationship between \ce{NaCl} and Diff2300 is robust with an $R^2$ value of 0.954; Diff800 can be used to estimate the level of  \ce{Pb(NO3)2} because the linear relationship between  \ce{Pb(NO3)2} and Diff800 is robust with an $R^2$ value of 0.920. 

\begin{table}[ht] 
  \caption{Regression performance $R^2$ for Diff2300 and Diff800.}
  \label{tab:freq}
  \begin{tabular}{ccl}
    \toprule
  Ingredient&Diff800&Diff2300\\
    \midrule
    \ce{Pb(NO3)2} &0.920 & 0.017\\
    \ce{NaCl} & 0.776 & 0.954\\
  \bottomrule
\end{tabular}
\label{table:tab2}
\end{table}

\subsection{Analysis on uncontrolled field samples}
\subsubsection{Composition of the field samples.} \label{ssec:composition}

\begin{figure*}
    \includegraphics[width=\linewidth]{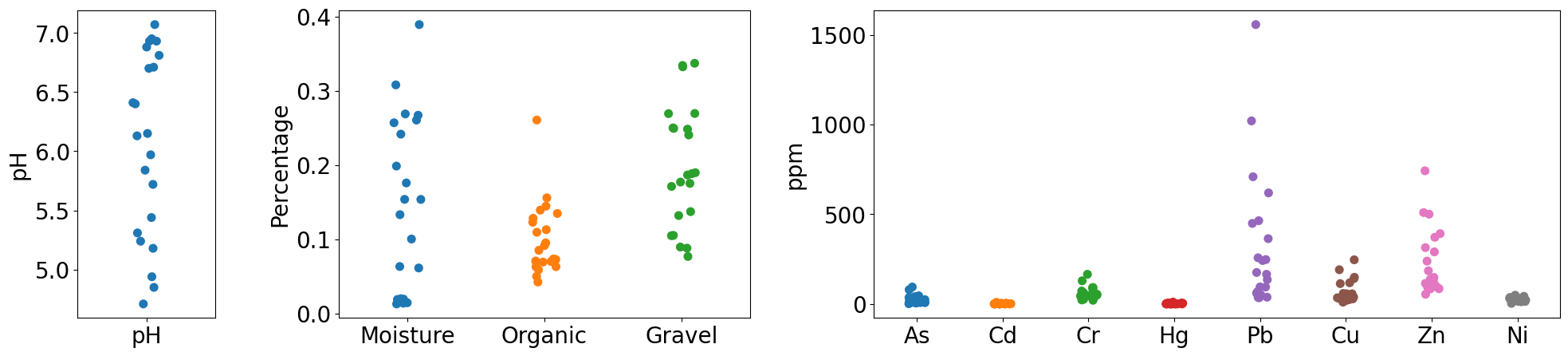}
    \caption{Composition of the field samples.}
    \label{fig: composition}
\end{figure*}

The studied field soil samples cover a wide range of moisture, pH, organic matter, gravel percentage, and heavy metals(including Pb). (Fig. \ref{fig: composition}). The average Pb content in those samples varied from 36ppm (minimum) to 1550ppm (maximum), with 1\revision{2} samples below and 1\revision{0} samples above the 200ppm threshold, moisture from $1\%$ to $40\%$, and organic from $4\%$ to $26\%$. The pH varied from 4.70 to 7.07, and the total salt amount from 58g to 710g. The composition of these samples is illustrated in Fig. \ref{fig: composition}.

\subsubsection{Model performance for the field samples}
The classification model's performance is evaluated by analyzing the confusion matrix. In this evaluation, we set model parameters $r=200$ and $k=100$.


\begin{table}
  \caption{Confusion matrix for classifying \ce{Pb} with the threshold of 200ppm.}
  \label{tab:conf_matrix_normalized}
  \centering
  \begin{tabular}{ccccc}
    \toprule
    & \multicolumn{2}{c}{Predicted Label} \\
    \cmidrule(lr){2-3}
    True Label & Pb: \textless 200 ppm  & Pb \textgreater= 200 ppm  \\
    \midrule
    Pb: \textless 200 ppm & 0.36 & 0.18 \\
    Pb: \textgreater= 200 ppm  & 0.091 & 0.36 \\
    \bottomrule
  \end{tabular}
\end{table}

\textbf{Confusion Matrix:}

Table \ref{tab:conf_matrix_normalized} is the confusion matrix that visually summarizes the performance of a classification model. It compares the actual labels of the data (ground truth) with the labels predicted by the model. Rows represent the actual labels(Pb content from pXRF measurement) in the data. Columns represent the labels predicted by the model. Parameters in the table include:

\textbf{True Negative (TN) - upper left: 36\%} where the model correctly predicted a negative class.

\textbf{True Positive (TP) - lower right: 36\%} where the model correctly predicted a positive class.

\textbf{False Positive (FP) - upper right: 18\%} where the model incorrectly predicted a positive class.

\textbf{False Negative (FN) - lower left: 9\%} where the model incorrectly predicted a negative class.

we analyze the confusion matrix by following metrics:

\textbf{Accuracy: 72\%}\revision{, t}he overall proportion of correct predictions $(TP + TN) / (Total)$\revision{.}


\textbf{Recall: 80\%}, defined as $(TP / (TP + FN)$\revision{,} the ratio of true positives (TP) to the sum of true positives and false negatives (TP + FN), is a crucial metric for evaluating classification models. In the context of \ce{Pb} detection, a high recall value is particularly significant. It indicates that the model effectively identifies a high proportion of samples with elevated \ce{Pb} levels, reducing the likelihood of overlooking critical cases. This characteristic strengthens the system's potential as a screening mechanism, prompting further investigation for potential \ce{Pb} contamination.


In addition, the false negative data points (which contain an excess amount of \ce{Pb} but are classified as less than 200ppm) do not contain any of the four samples higher than 500ppm of \ce{Pb}. We show that in our $22*k$ iterations \revision{(where $k$ is a parameter which indicates the number of times the ensemble model has been trained and tested)}, only the samples with 257ppm and 449ppm \revision{\ce{Pb}} were classified as false negatives. The low false negative rate of the model allows the system to be an excellent screening system for Pb detection. If our system's output is "above the threshold"\revision{,} the soil sample can be sent to a lab for further investigation.

\section{Limitations and Future Work}

SoilScanner demonstrates the feasibility of utilizing wireless-based technologies for detecting \ce{Pb} (and other salts or trace metals) in soil. Given the widespread presence of \ce{Pb} as a metal contaminant in urban soils, the capability to detect it holds significant implications for public health, urban agriculture, and sustainability. However, more work will be needed to achieve the goal of creating an affordable, accessible, portable, and non-invasive wireless sensor for in-situ \ce{Pb} detection in soil.

\subsection{Consideration of data collection and analysis approaches}

The SoilScanner was constructed and tested using two sets of samples: 23 controlled, lab-prepared soil for regression, and 22 uncontrolled, natural field samples for classification. These two approaches represent distinct strategies for addressing the detection problem.

\textbf{Controlled Sample Testing: Bottom-Up Approach}

In the regression model, the goal was to establish a mapping comparison between the predicted values of the system and the exact \ce{Pb} concentration with known methods (in this case, pXRF). This involved understanding the impact of various slats at different frequencies and determining the influence of soil composition. Calibration curves were developed for individual salts in this study \ce{NaCl} and \ce{Pb(NO3)2} in a controlled setting. However, field soils have greater complexity and variability in properties. Additional experiments are necessary to examine the effects of other soil components.

\textbf{Uncontrolled Sample Testing: Top-Down Approach}

In contrast, for uncontrolled field samples, we used a binary classification approach based on a threshold rather than mapping exact \ce{Pb} amounts. Acknowledging the variability in field soil samples, this approach considered the impact of \ce{Pb} on signal propagation alongside other factors. The model can be generalized and further developed to support multi-class classification by incorporating various samples with varying properties. This requires lab experiments with soil samples collected from different locations to serve as validation for the developed model. 



\textbf{Integration of Approaches: Future Directions}

While the two approaches necessitate different types of data collection and analysis, merging them offers a powerful strategy for future research. Leveraging the robust relationships established in controlled sample testing, this information can inform the binary classification model, reducing the complexity and sample size required for building an accurate and robust model. This integration of approaches represents a promising direction for advancing the capabilities of SoilScanner. \revision{We acknowledge the limited sample size in this study, and data augmentation, while helpful, is not a perfect substitute for real samples. Despite this, the experiment yielded promising results, which we expect to improve with greater accuracy and generalizability in future research. 
}

\subsection{Complex environmental variables}
The experiments in this study were conducted in a controlled lab setting. The application of the system to the real world requires consideration of other challenges, such as the presence of plants and other surface coverages, WiFi interference or objects that may affect reflection, as well as weather conditions that may affect RF propagation. 

\subsection{Portable chip manufacturing}
This study has yielded promising results regarding the use of commercial RFID devices. Similar to RFID setup, the use of existing WiFi chipsets in smart devices that are smaller in size and work on batteries \revision{will} be explored. If feasible, this hardware can then be put together to form a small sensor that is portable and can run on battery power. Research \cite{nandakumar20183d} has shown that it is possible to build a multiband RF system of size $15 mm^2$ that can operate on three frequency bands namely RFID (800MHz), WiFi (2.4 and 5GHz), and run on button cell battery.

\section{Conclusions}

In this paper, we present SoilScanner, a multiband  RF-based system for screening \ce{Pb} contamination in soil. SoilScanner utilized the mathematically proved assumption that the signal power spectrum carries the information of individual salts in the soil. We show the feasibility of separating \ce{Pb} in soil by building a regression model for controlled lab-prepared soil samples and a classification model for uncontrolled field soil samples based on the 200ppm threshold set by the U\revision{.}S\revision{.} EPA. Our results show that both models are robust, and the classification model has a zero error rate when \ce{Pb} is > 500ppm. We are confident that SoilScanner is a significant step toward building a low-cost, easy-to-use, and high-accuracy \ce{Pb} detection/screening system.


\section{Appendix}

\begin{figure}[hbtp]
    \centering
    \includegraphics[width=\linewidth]{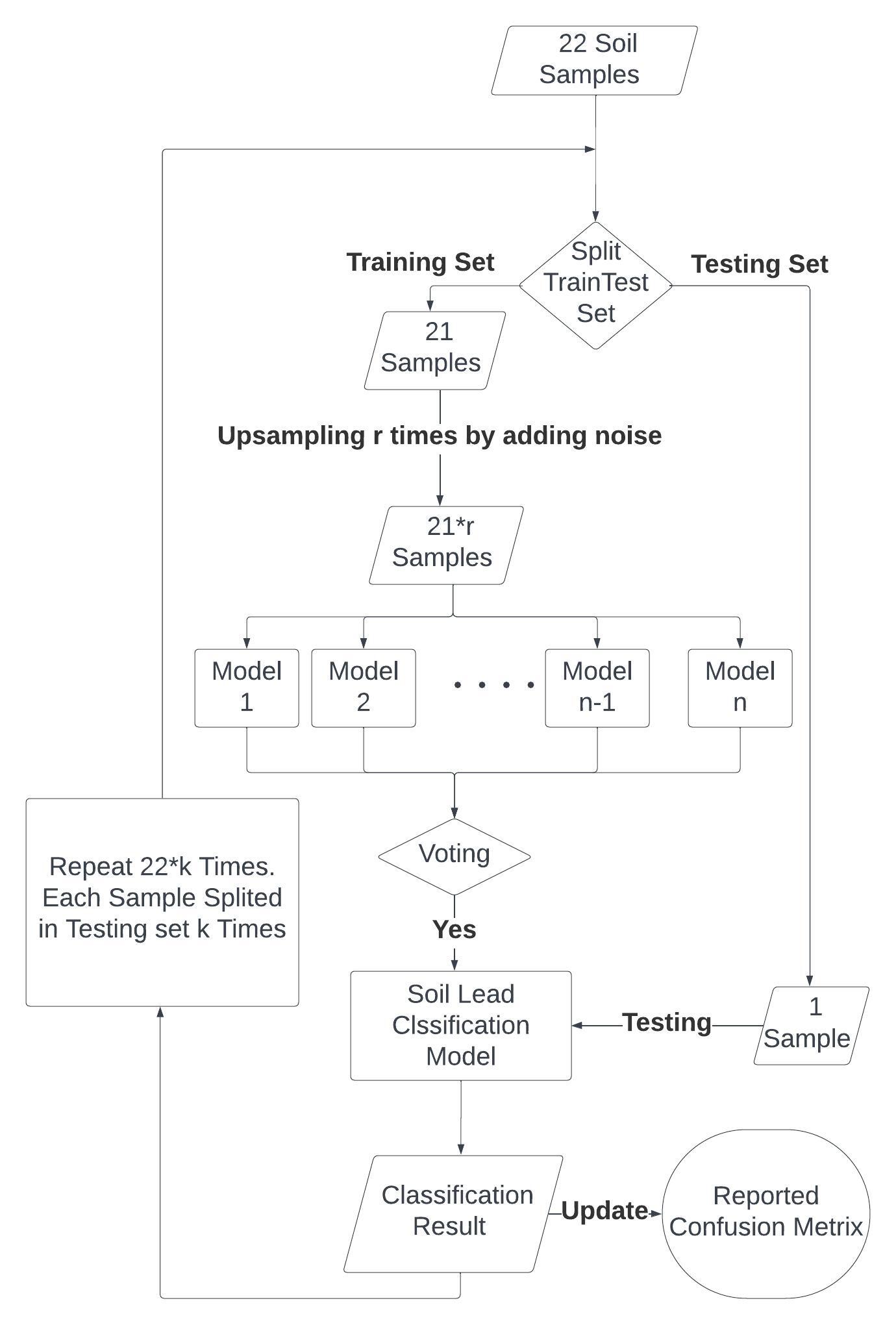}
    \caption{Pb classification system model.}
    \label{fig: Model_wild}
\end{figure}

The implementation details of the model designed for uncontrolled field samples are illustrated in Fig. \ref{fig: Model_wild}.

\begin{acks}
\revision{
We would like to thank Kaitlin McLaughlin, lab manager from the Urban Soils Lab at Brooklyn College, for her assistance in preparing and measuring the soil samples. This research was supported by a grant from the Cornell Institute for Digital Agriculture (CIDA).
}

\end{acks}

\bibliographystyle{ACM-Reference-Format}
\bibliography{refs}

\end{document}